\documentclass[10pt]{iopart}

\usepackage{iopams}
\usepackage{graphicx} 
\usepackage{dcolumn}
\usepackage{upgreek }
\usepackage{times}
\usepackage{textcomp}
\usepackage{ wasysym }
\usepackage{titlesec}
\usepackage{float}
\usepackage{color,xcolor}
\usepackage{graphicx,times}        
\usepackage{cuted} \stripsep -3pt plus 3pt minus 2pt
\usepackage{epstopdf}
\usepackage{graphicx}
\usepackage{subfigure}
\usepackage{makecell}
\usepackage{multirow}
\renewcommand{\figurename}{Figure}
\usepackage[colorlinks,linkcolor=blue, urlcolor=blue, anchorcolor=blue, citecolor=blue]{hyperref}
\usepackage{longtable,booktabs}
\usepackage{fouriernc}
\usepackage{epsfig,graphics,graphicx}
\usepackage{color}
\usepackage{flushend}
\usepackage{ulem}
\usepackage[sort&compress,square,numbers]{natbib}

\begin{document}
\title{Two-parameter Radial Equilibrium Models for Field-Reversed Configurations}

\author{
H. J. Ma$^{1,2,3}$, 
H. S. Xie$^{2,3*}$, 
Y. K. Bai$^{2,3}$, 
S. K. Cheng$^{2,3}$,
B. H. Deng$^{2,3}$, 
M. Tuszewski$^{4}$, 
Y. Li$^{2,3}$, 
H. Y. Zhao$^{2,3}$, 
B. Chen$^{2,3}$, 
and J.Y. Liu$^{1*}$
}

\address{
$^1$ Key Laboratory of Materials Modification by Laser, Ion and Electron Beams (Ministry of Education), School of Physics, Dalian University of Technology, Dalian 116024, People's Republic of China

$^2$ Hebei Key Laboratory of Compact Fusion, Langfang 065001, People’s Republic of China

$^3$ ENN Science and Technology Development Co., Ltd., Langfang 065001, People’s Republic of China

$^4$ ENN Consultant, Riverside, CA 92506, USA
}
\eads{*\mailto{xiehuasheng@enn.cn},  \mailto{jyliu@dlut.edu.cn}}

\begin{indented}
\item[\today]
\end{indented}

\begin{abstract}
A new equilibrium pressure profile extending the Rigid-Rotor (RR) model with a simple unified expression $P=P(\psi;\beta_{s},\alpha, \sigma)$ for both inside and outside the separatrix is proposed, in which the radial normalized field-reversed configuration (FRC) equilibrium profiles for pressure, magnetic field, and current can be determined by only two dimensionless parameters $\beta_s\equiv P_s/2\mu_0B_e^2$ and $\delta_s\equiv L_{ps}/R_s$, where $P_s$ is the thermal pressure at the separatrix, $B_e$ is the external magnetic field strength, $L_{ps}$ is the pressure profile scale length at the separatrix, and $R_s$ is the separatrix radius. 
This modified rigid rotor (MRR) model has sufficient flexibility to accommodate the narrow scrape of layer (SOL) width and hollow current density profiles, and can be used to fit experimental measurements satisfactorily.
Detailed one-dimensional (1D) characteristics of the new MRR model are investigated analytically and numerically, and the results are also confirmed in two-dimensional (2D) numerical equilibrium solutions. 
\end{abstract}
\submitto{\NF}
\maketitle
\ioptwocol

\section{Introduction}
Field-Reversed Configurations (FRC) are axisymmetric compact  toroids with no or weak toroidal magnetic field($B_\theta\simeq0)$\cite{armstrong1981field,finn1982field,Tuszewski1988Field,steinhauer2011review} . 
Because of its simple configuration, intrinsically high $\beta$, existence of natural divertor structure, good stability, and also the attractive feature of without material objects linking the torus which allows the FRC to be translated along the cylindrical axis and to be compressed in separate vacuum chambers\cite{Tuszewski1988Field,steinhauer2011review}, FRCs are often chosen as the advanced fuel fusion reactor design\cite{2010Dynamic} or as the target plasma for magnetized  target fusion (MTF) approach\cite{Tuszewski1988Field}. 
After formation, FRC plasmas are observed to be in stable equilibrium for many Alfv\'{e}n transit times\cite{spencer1985experimental}. 
The detailed features of FRC equilibrium is of great significance to the research of instability, transport, and compression processes.  

FRC is usually elongated cylindrical axisymmetric plasma, which can be well described in cylindrical coordinates  with ($r,\theta,z$). 
It is generally accepted\cite{armstrong1981field, finn1982field, Tuszewski1988Field, steinhauer2011review} that most FRC equilibria can be described by the Grad-Shafranov (G-S) equation. 
In cylindrical coordinates, with toroidal symmetry ($\partial{}/\partial{\theta}=0$), the G-S equilibrium equation is:
\begin{equation}
\Delta^*\psi\equiv r \frac{\partial}{\partial r}\left(\frac{1}{r} \frac{\partial \psi}{\partial r}\right)+\frac{\partial^{2} \psi}{\partial z^{2}}=-\mu_{0} r^2 P^{\prime}(\psi)  \label{eq:1} 
\end{equation}
with 
\begin{eqnarray}
J_{\theta}= r P^{\prime}(\psi) \label{eq:2}\\
B_z=\frac{1}{r} \frac{\partial{\psi}}{\partial{r}}; \quad B_r=-\frac{1}{r} \frac{\partial{\psi}}{\partial{z}} 
\end{eqnarray}
where $B_\theta=0$ is assumed, and $\psi=\Psi_p/2\pi=\int_{0}^{r} B_{z} r dr$ is the normalized radial magnetic flux, $J_{\theta}$ is the toroidal current density of plasma, $P^{\prime}$ is the derivative of pressure with respect to $\psi$ and $\mu_{0}$ is the vacuum permeability. 
Studies have shown that in present experiments kinetic effects of the equilibrium is small\cite{webster1991two}, and the FRC experimental study confirmed the applicability of the G-S equation\cite{armstrong1981field, okada1989reduction}, at least for the inner region of FRCs.

As is shown above, the pressure profile determines the FRC’s equilibrium. Various analytical pressure profile models, such as (extended) Hill's Vortex\cite{spencer1982free, steinhauer1990improved}, Rigid-Rotor (RR)\cite{morse1970rigid, armstrong1981field}, extended Solov'ev\cite{cerfon2010one}, 2PE\cite{steinhauer2009equilibrium}, and SYM\cite{lee2020generalized} models, have been proposed for theoretical and experimental purposes. 
In numerical simulations, dozens of pressure expressions for the numerical solution of G-S equation have been proposed\cite{Tuszewski1988Field, steinhauer2011review,kanno1995ideal,steinhauer2014two}. 
Among these models, the RR model is the most widely adopted. However, it has the disadvantage of profile stiffness (i.e., when the parameter changes the shape of the profile remains essentially the same), which can lead to excess SOL thickness, and sometimes it is difficult to fit the experimental profile data with the RR model.

Recently, a 3PE model is proposed by Steinhauer\cite{steinhauer2014two}, which is so far the most comprehensive $P(\psi)$ model for FRC equilibrium. 
However, the $P(\psi)$ form in the 3PE model is complicated and two piecewise expressions are used for the inner and outer regions, respectively,  making it inconvenient to use. 
In this article, a two-parameter Modified RR model (MRR model) is proposed. It is constructed based on the RR model with a simple unified (instead of piecewise) expression  $P=P(\psi; \beta_s, \alpha, \sigma)$, with the flexibility to allow for  hollow or peaked current profiles, and flexible SOL width to match experimental measurements.

The outline of this paper is as follows. 
Section \ref{sec:level2} introduces the formulation of the MRR model. 
The properties of the MRR model and the comparison of MRR with other models and experimental measurements are described in Section \ref{sec:level3}, where a new method for solving free parameters of the equilibrium model is also presented. 
FRC equilibrium in $(\beta_s,\delta_s)$ parameters space is explored in Section \ref{sec:level4}.
2D equilibrium based on the MRR model and solved using the GSEQ\cite{xie2019gseq} code is presented in Section \ref{sec:level5}.
Section \ref{sec:level6} concludes with a summary and outlook to future work.

\section{\label{sec:level2}THE MRR EQUILIBRIUM MODEL}
Although various models have been proposed to analyze the FRC experimental results, the RR model is the simplest one with clear physical meanings and is the most widely used\cite{deng2018first, conti2014rigid}. 
Therefore, the MRR pressure profile model $P=P(\psi)$ is based on the RR model, and is required to be as simple as possible, to have sufficient flexibility for fitting experimental data, and can be reduced to the RR model. 
In the following, the pressure profile in the RR model and the more recent 2PE and 3PE models are reviewed first, and then the pressure profile of the new MRR model is introduced.

The pressure expression of the RR model is
\begin{equation}
P=\frac{B_{e}^2}{2\mu_0} \cdot {sech}^{2}(K \cdot u) \label{eq:4}
\end{equation}
where $u\equiv 2r^2/R_{s}^2-1$ is the familiar minor radius variable, $R_s$ is the separatrix radius, $B_e$ is considered to be the vacuum magnetic field at the mid-plane, $K$ is a free parameter. From Ref.\cite{armstrong1981field}, the following relation between radial flux $\psi$ and $r$ is obtained:
\begin{equation}
\cosh(K \cdot u)=\exp \left(\frac{4 K}{B_{e} R_{s}^{2}} \psi\right) \cdot {cosh} K, \label{eq:5}
\end{equation}
From Eqs.\eref{eq:4} and \eref{eq:5},  the pressure profile in the RR model can be written as a flux function as in Eq.\eref{eq:6},
where $\beta\equiv P/(B_{e}^2/2\mu_{0})$ is defined as the ratio between thermal pressure and magnetic pressure, and $\beta_s$ is the separatrix value of $\beta$, with $\beta_s={sech}^{2}(K)$ in the RR model\cite{armstrong1981field}.

As is well known that with a single parameter, the profiles in the RR model is quite stiff, and the current profile is always singly peaked from the geometry center to the boundary [Figs.\ref{img_p2_r}-\ref{img_michel84} and Ref.\cite{steinhauer2014two}].
The SOL width is also much larger than some experimental measurements \cite{Tuszewski1988Field}.
Recently the more flexible 2PE and 3PE models are proposed.
The normalized $B_z$ inside of the separatrix of the 2PE\cite{steinhauer2008equilibrium} model is $b=b_{s} u e^{\sigma\left(u^{2}-1\right)}$ for $|u| \leq 1 $. 
We define the direction so that $\psi<0$ is inside of the separatrix (inner region), and $\psi>0$ in the outer region (SOL), and $\psi=0$ at the separatrix. 
The pressure profile $P(\psi)$ for the 2PE model is shown in Eq.\eref{eq:7}. 
The $P(\psi)$ for a more complicated 2PE model\cite{steinhauer2009equilibrium} with $b=b_{s} u e^{\sigma\left(u^{4}-1\right)}$ for $|u| \leq 1 $, is difficult to calculate analytically, thus not shown here.
The pressure profile $P(\psi)$ of the 3PE\cite{steinhauer2014two} model inside the separatrix is given in Eq.\eref{eq:8}.
In Eqs.\eref{eq:7} and \eref{eq:8} $b_s$ is the normalized magnetic field $b\equiv B(r)/B_e$ at the separatrix, $\sigma$ and $\alpha$ are free parameters.
In the 2PE and 3PE models [Eqs.\eref{eq:7} and \eref{eq:8}], the $\sigma$ term can control the current density profile in polynomial form, such as the term $\left(\frac{8 \psi \sigma}{B_{e} R_{s}^{2} b_{s}}+1\right)^{2}$ in 2PE model and the term $\frac{8 \psi}{B_{e} R_{s}^{2}} \cdot\left(1+\frac{1}{2} \frac{8 \sigma}{B_{e} R_{s}^{2}} \psi\right)$ in 3PE model.

\begin{strip}
\begin{eqnarray}
\fl
\mbox{RR:} \quad &P(\psi) =\frac{B_{e}^2}{2\mu_0} \cdot \exp \left(-\frac{8 K}{B_{e} R_{s}^{2}} \psi\right) \cdot {sech}^{2}(-K) =\frac{B_{e}^2}{2\mu_0} \cdot \beta_{s} \cdot \exp \left(-\frac{8 K}{B_{e} R_{s}^{2}} \psi\right)  \label{eq:6} \\
\fl \mbox{2PE($\psi \le0$):}  \quad  &P(\psi)=\frac{B_{e}^{2}}{2 \mu_{0}}\left\{1-b_{s}^{2}\left[\frac{1}{\sigma} \ln \left(\frac{8 \psi \sigma}{B_{e} R_{s}^{2}  b_{s}}+1\right)+1\right] \cdot\left(\frac{8 \psi \sigma}{B_{e} R_{s}^{2}  b_{s}}+1\right)^{2}\right\};\label{eq:7} \\
\fl \mbox{3PE($\psi \le0$):}  \quad  &P(\psi)=\frac{B_{e}^{2}}{2 \mu_{0}}\left\{\beta_{s}-\frac{8 \beta_{s} \alpha_{1}^{2}}{\left(7 \alpha_{1}-\sigma\right)} \left[ \frac{8 \psi}{B_{e}R_{s}^{2}} \left(1+\frac{1}{2} \frac{8 \sigma}{B_{e} R_{s}^2} \psi\right)+\left(\sigma+\alpha_{1}\right) \frac{1-\exp \left(\frac{8 \alpha_{2}}{B_{e} R_{s}^2}\ \psi\right)}{8 \alpha_{1}^{2}} \right] \right\};   \qquad \label{eq:8} \\
\fl \mbox{MRR:}  \quad  &P(\psi)=\frac{B_{e}^2}{2\mu_{0}} \beta_{s} \cdot \exp \left[-\alpha  \frac{\psi}{B_{e} R_{s}^{2}}  -f_{2}(\psi)\right] \cdot f_{1}(\psi)+f_{0}(\psi); \label{eq:9}\\
\fl \mbox{MRR-0:}  \quad  &P(\psi)=\frac{B_{e}^{2}}{2 \mu_{0}} \beta_{s} \cdot \exp \left(-\alpha \frac{\psi}{B_{e} R_{s}^{2}}   \right)+f_{0}(\frac{ \psi}{B_{e} R_{s}^{2}};\sigma)	; \label{eq:10}\\
\fl \mbox{MRR-1:}  \quad  &P(\psi)=\frac{B_{e}^{2}}{2 \mu_{0}} \beta_{s} \cdot \exp \left(-\alpha \frac{ \psi}{B_{e} R_{s}^{2}} \right) \cdot\left[\sigma \Big(\frac{ \psi}{B_{e} R_{s}^{2}} \Big)^q+1\right]^{n}, ~q=1, n=2 \mbox{ as default}; \label{eq:11}\\
\fl \mbox{MRR-2:}  \quad  &P(\psi)=\frac{B_{e}^{2}}{2 \mu_{0}} \beta_{s} \cdot \exp \left[-\alpha \frac{ \psi }{B_{e} R_{s}^{2}} -\sigma \Big(\frac{ \psi}{B_{e} R_{s}^{2}} \Big)^n \right], ~n=2  \mbox{ as default};\label{eq:12}
\end{eqnarray}
\end{strip}
The 2PE and 3PE models are piecewise defined and complicated, difficult to use in practical applications. 
Therefore the RR model is modified to support adjustable current density profiles and SOL width. As shown in Eq.\eref{eq:9}, new terms can be added to the RR model in three positions, denoted by $f_{0}$, $f_{1}$ and $f_{2}$ respectively. 
Although the functions $f_i(\psi)$ with $i=0,1,2$ can be arbitrary and coexist, we would consider the simplest case firstly, i.e., only one $f_i$ exists and $f_i$ is single parameter polynomial of $\psi$.
Thus, a series of Modified RR (MRR) models are constructed in Eqs.\eref{eq:10}-\eref{eq:12}. It will be shown later that these MRR models meet the requirements stated above. 
In these models, $B_e$ and $R_s$ are normalized scaling parameters and $\beta_s$, $\alpha$ and $\sigma$ are parameters to control the radial FRC profile shapes.
It should be noted that no strict relationship between $\alpha$ and $\beta_{s}$ is assumed in the MRR model in contrast to the RR model.
In order to distinguish different MRR models, we identify them as MRR-i for different $f_{i}$ with $i=0,1,2$.  
The 3PE model can also be regarded as a special but complicated case of MRR-0 model. 
Physically, $P \to 0$ is required far off the separatrix $(\psi\to\infty)$. However, it can be shown that $P(\psi\to\infty) \to \pm \infty$ in MRR-0 model, therefore another expression is needed in the open field line $\psi>0$ region. This is also why the 3PE model requires separate expressions for the inner and outer regions.
Hence in this work, only the MRR-1,2 models are studied in details. 
At the separatrix and geometry axis with $\psi=0$, the pressure $P=\beta_{s}B_{e}^2/2 \mu_{0} $. With $\sigma=0$, $\alpha=8K$ and $\beta_{s}=sech^2(K)$ in Eq.\eref{eq:11}-\eref{eq:12}, the MRR model reduces to the RR model.

The current profile index ($h$), which describes the distribution property of current inside the separatrix of FRC equilibrium, is introduced in Ref.\cite{steinhauer1992profile}:
\begin{equation}
h \equiv \frac{\left(j_{\theta} / r\right)_{O-point}}{\left\langle j_{\theta} / r\right\rangle} \label{eq:13}
\end{equation}
where the averaging is calculated inside the separatrix in the mid-plane\cite{kanno1995ideal, steinhauer2009equilibrium}:
\begin{equation}
\left\langle j_{\theta} / r\right\rangle \equiv \left(1 / \pi R_{s}^{2}\right) \int_{0}^{R_{s}}\left(j_{\theta} / r\right) 2 \pi r d r. \label{eq:14}
\end{equation}
The current density profile is “hollow” for $h<1$, i.e. hollow at the magnetic axis, “flat” for $h=1$, and “peaked” for $h>1$.
We will see in Figs.\ref{img_MRR1}- \ref{img_MRR21} that the current density profile of the MRR model can accommodate all three cases.

The following quantities would be useful in model applications:
\begin{eqnarray}
\left \langle \beta \right \rangle&=&\left(1 / \pi R_{s}^{2}\right) \int_{0}^{R_{s}}\beta 2 \pi r d r, \label{eq:15}\\
I_{in}&=& \int_{0}^{R_s}j_{\theta} d r, \label{eq:16}\\
I_{total}&=& \int_{0}^{\infty}j_{\theta} d r,\label{eq:17} \\
c_i&=&I_{in}/I_{total}, \label{eq:18}\\
c_p&=&\frac{\int_{0}^{R_s}p r dr}{\int_{0}^{\infty}p r dr}, \label{eq:19}
\end{eqnarray}
which are the average $\beta$, 
line current density inside the separatrix, 
total current density, 
the fractional current inside the separatrix, and the fractional pressure integral inside the separatrix, which represent the fractional number of particles inside the separatrix assuming constant temperature, respectively.
Using Eq.\eref{eq:16}, the total current inside the separatrix under racetrack  approximation can be calculated by multiplying the FRC length.

\section{\label{sec:level3}1D EQUILIBRIUM PROFILES OF THE MRR MODEL}
The RR model has only one free parameter, as shown in \ref{Append:A}, where a more general relationship of $K=\log \left[\sqrt{\beta_{s}}/ \left(\sqrt{1-\beta_{s}}+1\right)\right]$ is derived. For $\sigma=0$, the MRR model is reduced to the RR model. Therefore, although it appears that the MRR model has three free parameters, it actually has only two.
This is also true for the 3PE model. It is the constraint of the radial pressure balance that reduces one of the free parameter.
In the following, a simple step-by-step method to solve the quasi-1D equilibrium is proposed in section \ref{sec:level3-1}. 
In section \ref{sec:level3-2}, the MRR model is applied to compare with experimental data. 
In section \ref{sec:level3-3}, general characteristics of 1D equilibrium profiles from the MRR model are presented.

\subsection{\label{sec:level3-1}A step-by-step method for quasi-1D equilibrium solution}
Experimentally, $R_s$ and $B_e$ can be measured more easily than $\beta_s$ and $\delta_s$, where $\delta_s=L_{ps} /R_s$, with $L_{ps}=[P/(dP/dr)]_s$ being the pressure gradient-based SOL width, and the subscript `s' denotes the value in the mid-plane ($z=0$) and at the separatrix ($r=R_s$). 
Therefore, $\beta_s$ and $\delta_s$ are chosen to be the two free parameters in characterizing the 1D equilibrium.

The applicability of quasi-1D FRC equilibrium is discussed in \ref{Append:B}. 
In the quasi-1D equilibrium
\begin{equation}
P_{m}=P(\psi)+\frac{B_{z}^{2}}{2 \mu_{0}}=\frac{B_{e}^{2}}{2 \mu_{0}} \label{eq:20}
\end{equation}
which yields $B_{z}=\pm \sqrt{2 \mu_{0}\left[P_{m}-P(\psi)\right]}$.
With $d \psi =rB_{z} dr$, and using $B_{z}$, we have
\begin{eqnarray}
 \label{eq:21}
 \left\{
 \begin{array}{lcl}
-\frac{d \psi}{\sqrt{2 \mu_{0}\left[P_{m}-P(\psi)\right]}} &=r d r,  r \leq R_{0} \vspace{1ex}\\
\frac{d \psi}{\sqrt{2 \mu_{0}\left[P_{m}-P(\psi)\right]}} &=r d r,  r>R_{0}
\end{array}
\right.
\end{eqnarray}
we thus can readily prove that $R_s=\sqrt{2}R_0$, with $R_{0}$ is the O-point radius. 
The pressure gradient scale length $L_{p}$ is
\begin{equation}
L_{p}\equiv-\frac{P}{\frac{d P}{d r}}=-\frac{P(\psi)}{\frac{d P(\psi)}{d \psi} \cdot \frac{d \psi}{d r}}. \label{eq:22}
\end{equation}
From Eq.\eref{eq:21}, we obtain $d\psi /dr$ in the range of $r \leq R_{0}$
\begin{equation}
\frac{d \psi}{dr}=-r \sqrt{2 \mu_{0}\left[P_{m}-P(\psi)\right]}. \label{eq:23}
\end{equation}
At the O-point, the flux $\psi=\psi_{m}$ and pressure $P=P_{m}$, with $\psi_{m}$ being the normalized trapped flux, therefore
\begin{equation}
P(\psi_{m})=P_{m}.\label{eq:24}
\end{equation}
The dimensionless forms are used for the radius  $r/R_s\to r$,  the pressure $P/P_{m}\to p$, SOL width $L_{ps} /R_s \to \delta_s$, the flux $\psi / (B_e R_{s}^2) \to \psi$ and the current density $J_{\theta}/J_{m} \to J_{\theta}$, with $J_{m}=B_{e}/(2\mu_{0}R_{s})$.
Hereafter, unless otherwise stated, $\psi$ means dimensionless flux, $J_{\theta}$ means dimensionless current density, $r$ means dimensionless radius.
Respectively the Eqs.\eref{eq:22}, \eref{eq:24} and \eref{eq:21} are reduced to:
\begin{eqnarray}
&\delta_{s}=-\frac{p}{\frac{d p}{d r}}=-\frac{p(\psi)}{\frac{d p(\psi)}{d \psi} \cdot \frac{d \psi}{d r}}  \label{eq:25}\\
&p(\psi_{m})-1=0 	\label{eq:26}\\
&\int_{0}^{\psi_{m}} \frac{d \psi}{\sqrt{\left[1-p(\psi_{m})\right]}}+\frac{1}{4}=0. \label{eq:27}
\end{eqnarray}

In terms of the normalized parameters defined above, Eq.\eref{eq:25} can be written in different forms for each model described by Eqs.\eref{eq:6}-\eref{eq:12}:
\begin{eqnarray}
\mbox{RR: } &8\delta_{s} \log \left(\frac{\sqrt{1-\beta_{s}}+1}{\sqrt{\beta_{s}}}\right) \sqrt{1-\beta_{s}}-1=0 \label{eq:28} \\
\mbox{MRR-1: } &\left\{\begin{array}{lcl}
\delta_{s}(\alpha-n \sigma) \sqrt{1-\beta_{s}}-1=0; &q=1 \vspace{1ex} \\
\delta_{s} \alpha \sqrt{1-\beta_{s}}-1=0; &q \neq 1
\end{array}\right. \label{eq:29} \\
\mbox{MRR-2: } &\delta_{s} \alpha \sqrt{1-\beta_{s}}-1=0 \label{eq:30} \\
\mbox{2PE: }  &8\delta_{s} (1-\beta_{s})(1+2\sigma)-\beta_{s}=0\label{eq:31} \\
\mbox{3PE: } &\delta_{s} 16 \alpha\left(3 \alpha-\sigma\right) \sqrt{1-\beta_{s}}+\left(7 \alpha-\sigma\right)=0 \label{eq:32}
\end{eqnarray}
The following steps are used to solve the constraint equations: First, the $\alpha$ parameter is solved from Eqs.\eref{eq:28}-\eref{eq:32} from specified $\delta_s$ and $\beta_s$; second, the intermediate variable of trapped flux $\psi_{m}$ is solved from Eq.\eref{eq:26}; third,  Eq.\eref{eq:27} is solved as $F(\sigma)=0$ to obtain $\sigma$; and finally all the other parameters are obtained and $P(\psi)$ is determined. Subsequently the FRC profiles such as $p(r)$, $B_z(r)$, $\psi(r)$ and $j_\theta(r)$ are determined.

\subsection{\label{sec:level3-2}Model comparison by fitting to the experimental data}
From the above discussions, it is shown that equilibrium profiles can be determined by a set of parameters ($\beta_s$,$\delta_s$). 
To compare different models, they are used to fit the experimental data from FRC experiments of NUCTE-III\cite{ohkuma1998production,ikeyama2009beta} and FRX-C\cite{tuszewski1984experimental}.

\begin{figure}[htbp]
\includegraphics[width=8cm]{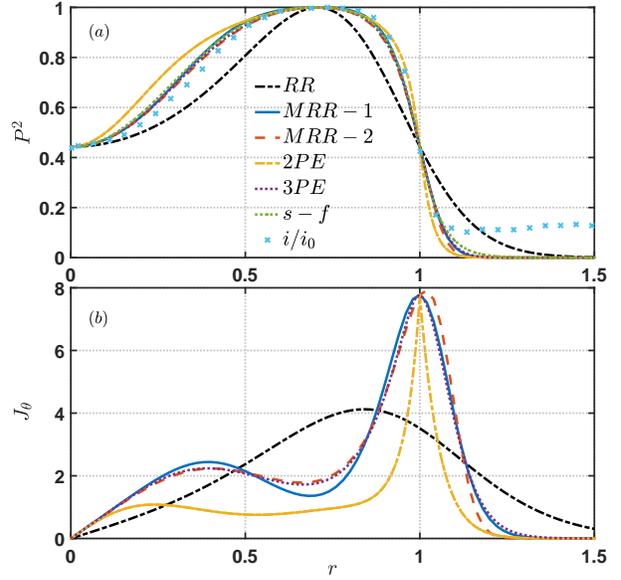}
\caption{The radial profiles of the RR, MRR-1,2, 2PE, 3PE and s-f models for NUCTE-III experimental parameters $(\beta_s=0.6655,\delta_s=0.148)$. (a) is the radial profile of bremsstrahlung intensity $i(r)/i_{0}$ from NUCTE-III\cite{ikeyama2009beta} and the squares of pressures $P^2/P_{m}^2$ of models, (b) is the toroidal current density. $K=0.66$ and $R_s=5.61cm$ is set. }
\label{img_p2_r}
\end{figure}

The normalized pressure profile in NUCTE-III device is represented by the measured normalized bremsstrahlung intensity profile as shown in Fig.\ref{img_p2_r}a by the crosses.
Bremsstrahlung radiation intensity is proportional to the square of density, and is a complicated function of $T_e$. In FRC plasmas, $T_e$ profile is quite flat within the separatrix\cite{Tuszewski1988Field,deng2012electron}. 
Therefore, the normalized bremsstrahlung intensity profile can be considered to be proportional to the square of the pressure.
In Fig.\ref{img_p2_r}a, quasi-1D equilibrium pressure squared from each model that best fit the experimental data are plotted.
It is shown that the RR model cannot fit the experimental data with broad peak.
This shows the disadvantage of profile stiffness of the RR model.
 It is also shown that within $r<0.5$, the curve fit of the 2PE model deviates from the data more than the MRR models and the 3PE model.
 The difference between the MRR models and the 3PE model is small. However, the MRR models have unified expression for both inside and outside the separatrix and are more convenient for applications.
 It should be noted that the 2PE, 3PE and s-f profile are given by segmented expressions presented in Refs.\cite{steinhauer2008equilibrium,steinhauer2014two,ikeyama2009beta}.
 The s-f profile is just proposed to fit the the experimental measurement in Ref.\cite{ikeyama2009beta}, meanwhile, the profile is a segmented expressions of r, which is not applicable to the G-S equation, so we will not study it further in the following.
 Fig.\ref{img_p2_r}b compares the current density profiles of difference models. 
 The RR model shows a  single peak, which is inconsistent with the broad experimental pressure profile shown in Fig.\ref{img_p2_r}a.
 The MRR models and the 3PE model yield similar hollow current profiles, while the 2PE model yields a much sharper current density peak near the separatrix.
 In equilibrium the pressure gradient is maintained by the $J\times B$ force.
 The broad pressure profile indicates the existence of a gradient minimum and is consistent with the hollow current density profiles from the MRR and 3PE models.

\begin{figure}[htbp]
\centering
\includegraphics[width=8cm]{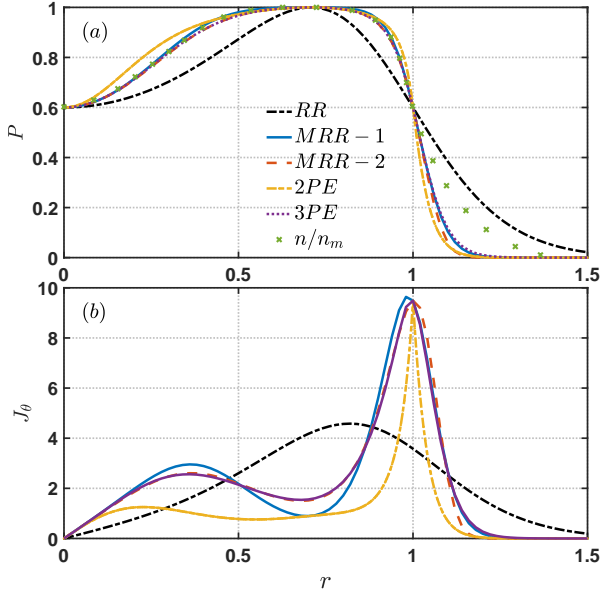}
\caption{The radial profiles of the RR, MRR-1,2, 2PE and 3PE models for FRX-C experimental parameters $(\beta_s=0.60,\delta_s=0.10)$.  (a) is  pressure profile from FRX-C experiment\cite{tuszewski1984experimental} with the constant temperature and the matched pressure profile of models. (b) is the toroidal current density.}
\label{img_michel84}
\end{figure}

The normalized experimental density profile of the FRX-C device from Ref.\cite{tuszewski1984experimental} is reproduced in Fig.\ref{img_michel84}a, represented by the crosses.
Assuming uniform electron temperature profile inside the separatrix, the shape of the pressure profile is close to the density profile.
Therefore, the pressure profile of each model can be used to fit the measured density profile as shown in Fig.\ref{img_michel84}a.
It is shown that the MRR and 3PE models fit the experimental data well inside the separatrix while the fit of the 2PE and RR models are not satisfactory.
The current density profiles of difference models is also presented in Fig.\ref{img_michel84}b. Like Fig.\ref{img_p2_r}b, the RR model shows a  single peak, and the 2PE model shows a great difference with the MRR and 3PE models, which displays the consistent results with Fig.\ref{img_michel84}a.

The MRR-1,2 and 3PE models also agree with each other well outside the separatrix, but are obviously different from the experimental density profile.
One possible reason is that the temperature profile is not constant but decreasing rapidly outside the separatrix\cite{deng2012electron}, which causes the pressure profile to drop faster than the density profile.
We have also checked with other experimental results, for examples the density profiles from Refs.\cite{okada1989reduction, suzuki1986numerical}, and similar conclusions can be obtained.
More restricted comparison between measured FRC profiles and the models should be done when more experimental data become available.

\subsection{\label{sec:level3-3}1D characteristics of MRR models}

From Figs.\ref{img_p2_r} and \ref{img_michel84}, it is found that the 3PE model is very close to MRR-2 model in the inner region and close to MRR-1 model at the outer region. So, the 3PE model is between MRR-1 and MRR-2 models. Thus in this subsection, only the properties of MRR-1 and MRR-2 models are studied in details.

\begin{figure}[htbp]
\centering
\includegraphics[width=8cm]{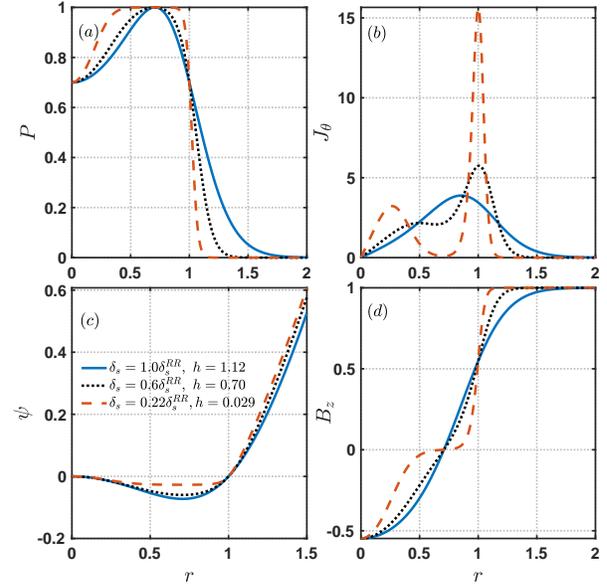}
\caption{The radial profiles of MRR-1 model using $n=2,q=1$ with ($\beta_{s}=0.7$,$\delta_{s}^{RR}=0.37$). (a) is the pressure $P$, (b) is the current density $J_{\theta}$, (c) is the magnetic flux $\psi$ and  (d) is the magnetic field $B_z$. $\delta_{s}^{RR}$ is the RR model's $\delta_s$, $h$ is the definition of Eq.\ref{eq:10}.}
\label{img_MRR1}
\end{figure}

\begin{figure}[htbp]
\centering
\includegraphics[width=8cm]{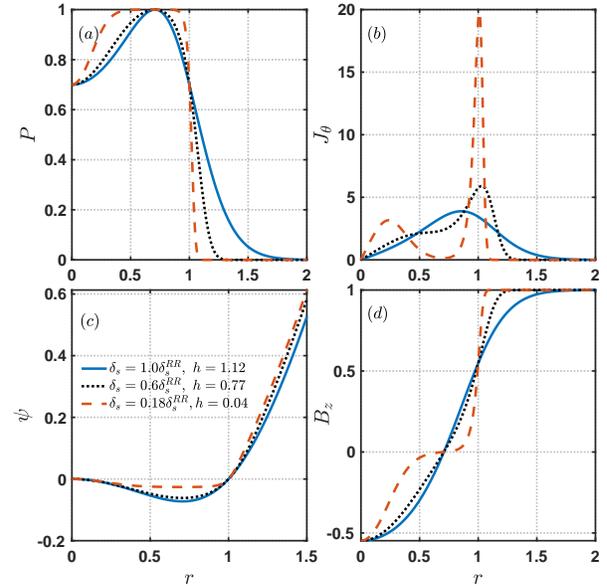}
\caption{The radial profiles of MRR-2 model using $n=2$ with ($\beta_{s}=0.7$,$\delta_{s}^{RR}=0.37$). (a) is the pressure $P$, (b) is the current density $J_{\theta}$, (c) is the magnetic flux $\psi$,  and  (d) is the magnetic field $B_z$. $\delta_{s}^{RR}$ is the RR model's $\delta_s$, $h$ is the definition of Eq.\ref{eq:10}.}
\label{img_MRR2}
\end{figure}

Fig.\ref{img_MRR1} shows the evolution of the radial profiles with the change of $\delta_s$ of MRR-1 model when $\beta_{s}$ is kept constant.
As shown in Fig.\ref{img_MRR1}a, with the decreases of $\delta_s$, the pressure profile become sharper in the outer region and flatter in the inner region of the FRC. The current density profile gradually changes from peak to hollow as shown in Fig.\ref{img_MRR1}b. 
The changes in magnetic flux $\psi$ and magnetic field $B_z$ are shown in Fig.\ref{img_MRR1}c and \ref{img_MRR1}d respectively.
These results show that when the current profile peaks sharply at the separatrix, the pressure gradient is sharp there, while the pressure peak is broad inside the separatrix, corresponding to the hollow current profile inside, which can be understood via the relation $J_\theta=rP^{\prime}(\psi)$. 
Similar results are obtained from the MRR-2 model with the variation of $\delta_s$ as shown in Fig.\ref{img_MRR2}.

\begin{figure}[htbp]
\centering
\includegraphics[width=8cm]{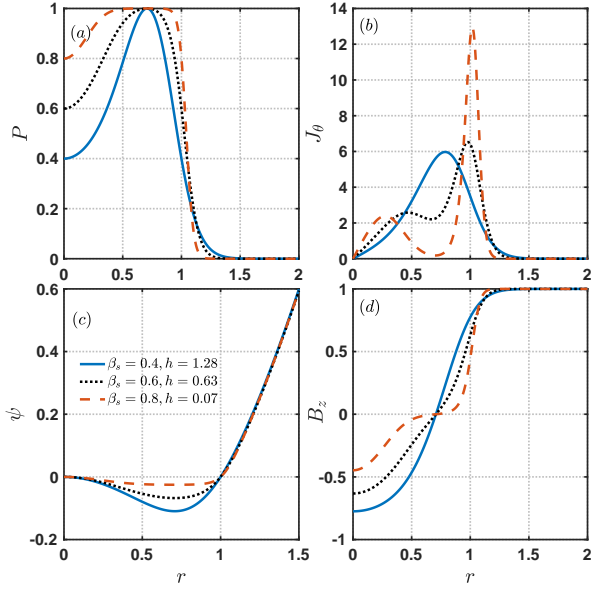}
\caption{The radial profiles of MRR-1 model using $n=2,q=1$ with $\delta=0.148$. (a) is the pressure $P$, (b) is the current density $J_{\theta}$, (c) is the magnetic flux $\psi$ and  (d) is the magnetic field $B_z$. }
\label{img_MRR11}
\end{figure}

\begin{figure}[htbp]
\centering
\includegraphics[width=8cm]{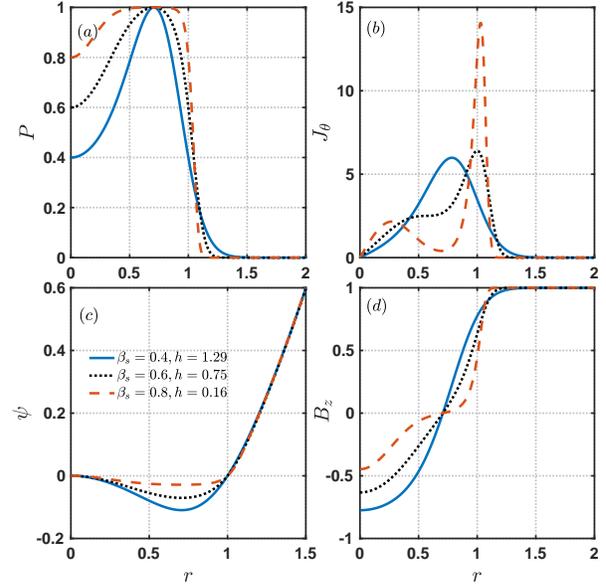}
\caption{The radial profiles of MRR-2 model using $n=2$ with $\delta=0.148$. (a) is the pressure $P$, (b) is the current density $J_{\theta}$, (c) is the magnetic flux $\psi$  and  (d) is the magnetic field $B_z$.}
\label{img_MRR21}
\end{figure}

Fig.\ref{img_MRR11} shows the evolution of the radial profiles with $\beta_s$ of MRR-1 model for constant $\delta_{s}$. 
As shown in Fig.\ref{img_MRR11}a, with the decreases of $\beta_s$, the pressure profile becomes flatter in the outer region and sharper in the inner region of the FRC. 
The current density profile gradually changes from hollow to peak is shown in Fig.\ref{img_MRR11}b. 
The changes in magnetic flux $\psi$ and the magnetic field $B_z$ are also presented in Fig.\ref{img_MRR11}c and \ref{img_MRR11}d respectively.
Similar results are obtained from the MRR-2 model with the variation of $\beta_s$ as shown in Fig.\ref{img_MRR21}.

We also observed in Figs.\ref{img_MRR1}- \ref{img_MRR21} that the differences between the radial profiles of MRR-1 and MRR-2 are very small for a wide range of $\beta_{s}$ and $\delta_s$. 
Thus again, the MRR models are robust. 
This implies that, instead of the form of $P(\psi)$, the parameters $(\beta_s,\delta_s)$ is much important to determine the FRC radial profiles, which is the most interesting finding in this work. 
Hence, we think the two parameters $(\beta_s,\delta_s)$ can describe FRC radial equilibrium completely to a very high accuracy. 
That is, after two values $(\beta_s,\delta_s)$ is given, a FRC equilibrium with profiles $\psi(r)$, $p(r)$, $J_\theta(r)$ and $B_z(r)$ is uniquely determined! 
This feature would simplify the FRC equilibrium reconstruction or fitting a lot.

\section{\label{sec:level4}FRC EQUILIBRIUM IN ($\beta_{s}, \delta_s$) PARAMETERS SPACE}
\begin{figure*}[htbp]
\centering
\includegraphics[width=16cm]{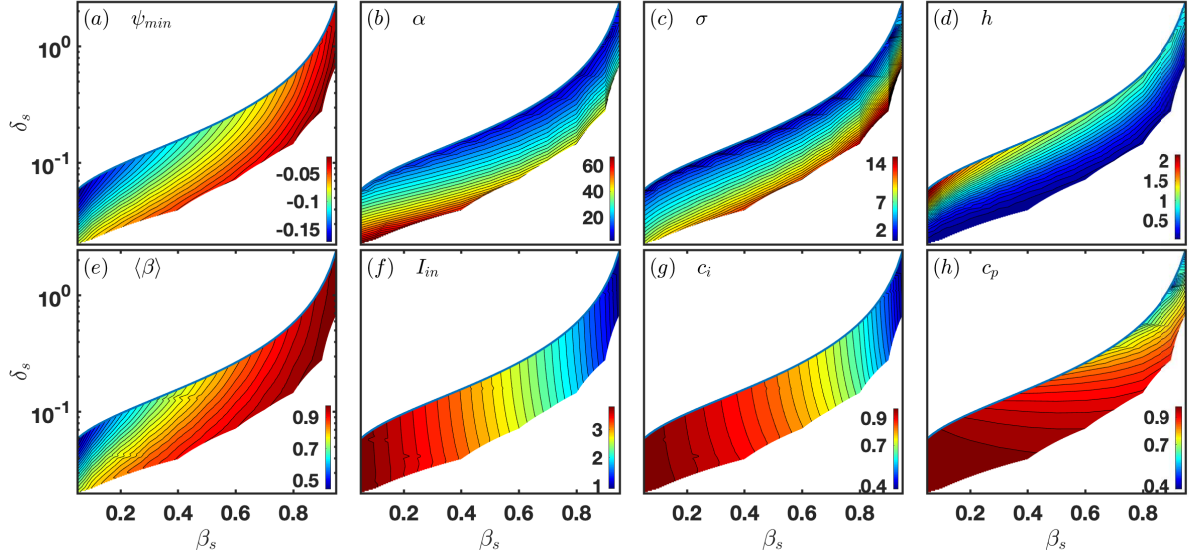}
\caption{The contour of other quantities in ($\beta_{s}, \delta_s$) space are shown. Trapped flux $\psi_{min}$ is (a), the parameters $\alpha$ is (b), the parameters $\sigma$ is (c), the current profile index $h$ is (d), the average $\beta$ $\left \langle \beta \right \rangle$ is (e), the line current density inside the separatrix with Eq.\ref{eq:16} is (f), the ratio of current inside the separatrix with Eq.\ref{eq:18} is (g) and the ratio of pressure inside the separatrix with Eq.\ref{eq:19} is (h).}
\label{img_MRR1-contour}
\end{figure*}

It is shown in \Sref{sec:level3} that a set of parameters ($\beta_{s}, \delta_s$) can represent FRC radial equilibrium uniquely for models with less than three free parameters, it would be useful to calculate other model parameters in the ($\beta_{s}, \delta_s$) parameters space. 
It is limited to the MRR-1 model in this section. 
These results can be used for future theoretical or experimental studies directly. 
Fig.\ref{img_MRR1-contour} shows the contour of quantities in the ($\beta_s$,$\delta_s$) space  using the MRR-1 model. 
The upper limit in the ($\beta_s$,$\delta_s$) space is the RR model and the lower line corresponds to $h=0$, $h$ is defined in Eq.\ref{eq:13}.
In FRC experiments, real-time rapid reconstruction of equilibrium is desired, after the values of ($\beta_s$,$\delta_s$) is given in the experiment, the equilibrium quantities including the trapped flux ($\psi_{min}$), the current profile index ($h$), the average $\beta$ ($\left \langle \beta \right \rangle$) can be estimated quickly using Figs.\ref{img_MRR1-contour}a, \ref{img_MRR1-contour}d, and \ref{img_MRR1-contour}e respectively, the current of plasma inside the separatrix, the ratio of current and the number of particles inside the separatrix are also given in Figs.\ref{img_MRR1-contour}f, \ref{img_MRR1-contour}g and \ref{img_MRR1-contour}h. 
Meanwhile, Figs.\ref{img_MRR1-contour}b and \ref{img_MRR1-contour}c can provide a set of initial free parameters ($\alpha$,$\sigma$) for accurate FRC equilibrium reconstruction.

\section{\label{sec:level5}MRR MODEL APPLIED TO 2D EQUILIBRIUM STUDIES}
\begin{figure}[htbp]
\centering
\includegraphics[width=8cm]{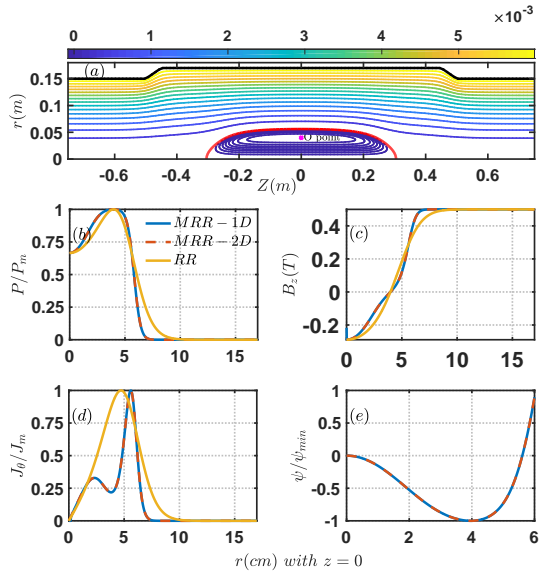}
\caption{The 2D contour of $\psi$ for NUCTE-III device and the comparison of the 1D and 2D radial profiles of MRR-1 model. (a) is the 2D contour of $\psi$, (b) is the radial pressure $P/P_{m}$, and (c) is the $B_{z}$ profile, (d) is current density $J_{\theta}/J_{m}$ profile at the mid-plane and (e) is $\psi$.}
\label{img_MRR-2D}
\end{figure}

The GSEQ (General Solver for EQuilibrium) code is developed by the  ENN theory and simulation group to study 2D FRC equilibrium properties by solving the G-S equation\cite{xie2019gseq} under fixed or free boundary conditions. 
Conducting cylindrical fixed boundary is used as upper wall conditions in the current version. 
The detailed description of the code and more 2D studies of the MRR-1,2 model will be presented in future publications. 
Here 2D FRC equilibrium with the MRR-1 model is solved by the GSEQ code and compared with the quais-1D result in Fig.\ref{img_MRR-2D}.
The MRR-1 model parameters chosen are $q=1, n=2$.
The NUCTE-III device geometric parameters are used.
It should be noted that the 2D result of MRR-1 is obtained using the same free parameters as in the 1D model with only the iteration of $\psi$.
From Fig.\ref{img_MRR-2D}b-d, it is seen that the 1D and 2D results of the MRR-1 model agree with each other well.
Similar comparison shows that the 1D and 2D results of the MRR-2 model also agrees well.
This means that the MRR-1,2 model is suitable for 2D FRC equilibrium reconstruction and the quasi-1D equilibrium holds well in the mid-plane.

\section{\label{sec:level6}SUMMARY AND CONCLUSION}
A modified rigid rotor FRC equilibrium pressure profile model $P(\psi;\beta_s,\alpha,\sigma)$ with unified simple expression for both inside and outside the separatrix is developed.
With two free parameters, which can be conveniently chosen to be the normalized SOL width and the separatrix $\beta$ value, this model eased the difficulty of profile stiffness of the RR model.
The MRR model is easy to solve with a simple step-by-step method, and is used to study the 1D and 2D FRC equilibrium properties.
This model shows its advantages in fitting the experimental profiles with broad pressure peak and sharp gradient near the separatrix.
In the future, FRC equilibrium fitting (FRC-EFIT) code can be developed based on the MRR model. 
Also, a more accurate study of some analytical approximated axial pressure equilibrium relations such as the the Barnes average $\beta$ relation\cite{barnes1979compact} and the trapped flux relation\cite{Tuszewski1988Field} will be a future topic.

\ack
This work is supposed by the China central government guides the development of local science and technology funding No.206Z4501G and the compact fusion project of the ENN group. 

\appendix{}
\section{\label{Append:A}THE RR MODEL HAS ONLY ONE FREE PARAMETER}
The  normalized pressure as a flux function $\psi$ in RR model is\cite{armstrong1981field}:
 \begin{equation}
p(\psi)=\beta_{s} \cdot \exp \left(-K \psi\right); \quad  \beta_{s}=sech^{2}(K)  \label{AC-1}
\end{equation}
where the pressure is normalized by the external magnetic pressure $B_e^2/2\mu_0$, the flux is normalized by $B_eR_s^2$.
We show in the following that the relationship between $\beta_s$ and $K$ is a consequence of radial pressure balance, and therefore, the RR model has only one free parameter.

Consider a more general normalized expression:
\begin{equation}
p(\psi)=\beta_s \exp (-b \psi) \label{AC-2}
\end{equation}
with two free parameters $\beta_s$ and $b$. 

Eq.\eref{AC-3} is obtained from radial pressure balance\cite{armstrong1981field}:
\begin{equation}
\frac{d \psi}{-\sqrt{\left[1-\beta_s \exp (-b \psi)\right]}}=\frac{r^2}{2}, \quad 0 \leq r \leq R_{0} \label{AC-3}
\end{equation} 
Integrate Eq.\eref{AC-3}, with the integration formula:
 \begin{equation}
\int \frac{1}{\sqrt{1-m e^{-n x}}} d x=\frac{2 \log \left[\sqrt{e^{n x}-m}+e^{\frac{n x}{2}}\right]}{n}  \label{AC-4}
\end{equation}
it can be obtained  
\begin{equation}
\eqalign{\sqrt{e^{b \psi}-\beta_{s}}+\exp\left(\frac{b \psi}{2}\right)   \cr
=\exp \left[-\frac{r^{2}}{4} b +\log [\sqrt{1-\beta_{s}}+1]\right]}  \label{AC-5}
\end{equation}
Setting $y \equiv \exp \left( b \psi /2 \right)$ and 
\begin{equation}
c \equiv \exp \left[-\frac{r^{2}}{4} b +\log [\sqrt{1-\beta_{s}}+1]\right]	\label{AC-6}
\end{equation}
the Eq.\eref{AC-5} is simplified to
\begin{equation}
\sqrt{y^{2}-\beta_{s}}=c-y. \label{AC-7}
\end{equation}
Solving Eq.\eref{AC-7} to obtain $y=(c^{2}+\beta_{s})/2 c$, thus
\begin{equation}
\psi=\frac{2}{b} \log \left\{\frac{c^{2}+\beta_{s}}{2 c}\right\}. \label{AC-8}
\end{equation}
At the O point with $r_0=1/ \sqrt{2}$, $P(\psi)=1$, then
\begin{equation}
\left\{\frac{c^{2}+\beta_{s}}{2 c}\right\}^{-2}=\frac{1}{\beta_{s}}. \label{AC-9}
\end{equation}
So $c^2=\beta_{s}$, which is 
\begin{equation}
\left[\sqrt{1-\beta_{s}}+1\right]^{2} \exp \left[-\frac{r^{2}}{2} b \right]=\beta_{s}. \label{AC-10}
\end{equation}
Then, we obtain
\begin{equation}
 b=-\frac{2}{r_{0}^{2}} \log \left(\frac{\beta_{s}}{[\sqrt{1-\beta_{s}}+1]^{2}}\right). \label{AC-11}
\end{equation}
Define $d \equiv c/ \sqrt{\beta_{s}}$, Eq.\eref{AC-2} can be written as
\begin{equation}
\eqalign{P(r)=\beta_s \exp \left(-2 \log \left\{\frac{c^{2}+\beta_{s}}{2 c}\right\}\right)=\left\{\frac{2}{d+d^{-1}}\right\}^{2} \cr
={sech}^{2}\left[\log \left(\frac{\sqrt{\beta_{s}}}{[\sqrt{1-\beta_{s}}+1]}\right)\left(2r^2-1\right)\right]} \label{AC-13}
\end{equation}
with 
\begin{equation}
d=\exp \left[\log \left(\frac{\sqrt{\beta_{s}}}{[\sqrt{1-\beta_{s}}+1]}\right)\left(2 r^2-1\right)\right] \label{AC-14}
\end{equation}
From Eq.\eref{AC-13}, we conclude that the general pressure expression $p=\beta_s\exp(-b\psi)$ has only one free parameter and is actually identical to the RR model.

\section{\label{Append:B} THE ACCURACY OF QUASI-1D EQUILIBRIUM}
\begin{figure}[htbp]
\centering
\includegraphics[width=0.47\textwidth]{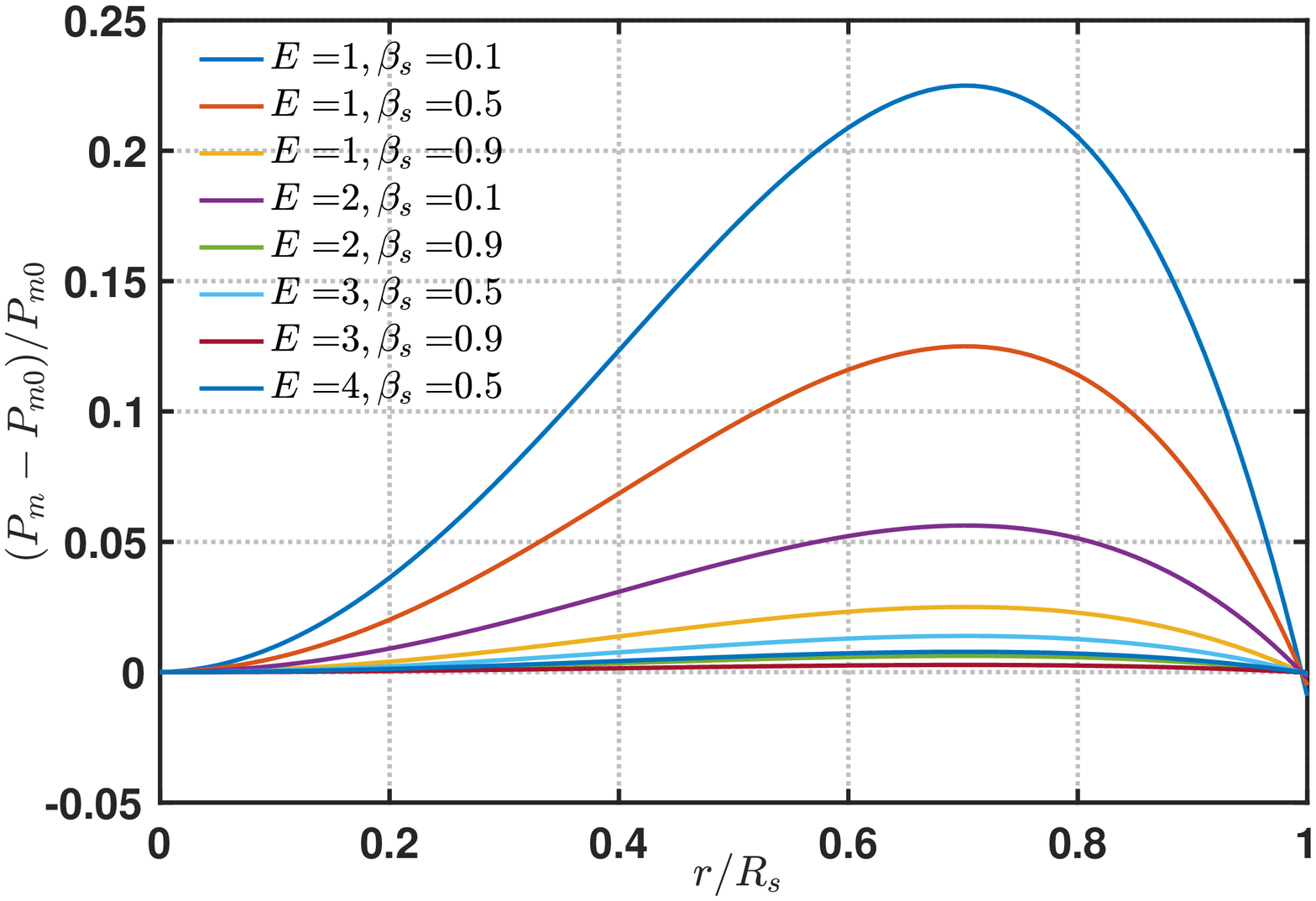} 
\caption{The effect of FRC elongation with Hill's model, which represents the accuracy of quasi-1D equilibrium.}
\label{img_hill}
\end{figure}

FRC is elongated along the cylindrical axis, and often considered as in 1D equilibrium when the curvature force at the mid-plane is neglected, leading to the following ``pressure balance" equation:
\begin{equation}
P+\frac{B_{z}^{2}}{2 \mu_{0}}=\frac{B_{e}^{2}}{2 \mu_{0}} \label{AD-1}
\end{equation} 
It is interesting to check the accuracy of the quasi-1D equilibrium.
The force balance equation without approximation is:
\begin{eqnarray}
&J \times B=\nabla P, \label{AD-2} 
\end{eqnarray}
In cylindrical coordinates ($r, \theta,z$), it can be written as:
\begin{equation}
J_{\theta} B_{z}-J_{z} B_{\theta}=\frac{\partial P}{\partial r}.  \label{AD-3} 
\end{equation}
For ideal FRC with $B_\theta=0$, using Ampere’s law, $ \mu_{0} J=\nabla \times \boldsymbol{B}$, Eq.\eref{AD-3} is integrated to yield:
\begin{equation}
P+\frac{B_{z}^{2}}{2 \mu_{0}}=\frac{B_{e}^{2}}{2 \mu_{0}}+\int \frac{B_{z}}{\mu_{0}}\left(\frac{\partial B_{r}}{\partial z}\right) d r \label{AD-4}
\end{equation}
where $B_e$ is the external magnetic field, and the last term representing the curvature effect is neglected in the quasi-1D equilibrium. 
The significance of this term is estimated below.
Define:
\begin{eqnarray}
\delta  \equiv \frac{\int \frac{B_{z}}{\mu_{0}}\left(\frac{\partial B_{r}}{\partial z}\right) d r}{\frac{B_{e}^{2}}{2 \mu_{0}}} 
=\frac{2}{B_{e}^{2}} \int B_{z}\left(\frac{\partial B_{r}}{\partial z}\right) d r. \label{AD-7}
\end{eqnarray}
In experiments the FRC separatrix is elliptical in shape. To estimate the value range of $\delta$, Hill’s vortex model with elliptical separatrix is used, where the flux contour is
\begin{equation}
\psi=\frac{B_{0}}{2} r^{2}\left(1-\frac{r^{2}}{a^{2}}-\frac{z^{2}}{b^{2}}\right) \label{AD-8}
\end{equation}
where $B_{0}=B_e\sqrt{1-\beta_s}$ is the magnetic field at the separatrix. in the mid-plane,
\begin{equation}
\delta=\frac{B_{0}^{2}}{B_{e}^{2} b^{2}} r^{2}\left(1-\frac{r^{2}}{a^{2}}\right). \label{AD-9}
\end{equation}
The maximum value of $\delta$ can be expressed in term of FRC elongation $E=b/a$:
\begin{equation}
\delta<\delta_{c}=\frac{1-\beta_{s}}{4 E^{2}} \label{AD-10}
\end{equation}
In current FRC experiments, typical $E$ value range is $3-10$. From Eq.\eref{AD-10}, when $E>2$, $\delta<1/16$, which means that the magnetic field curvature has little effect on the radial balance and quasi-1D equilibrium is a good approximation for FRC plasmas.
Fig.\ref{img_hill} plotted more detailed estimation of $\delta$ for typical FRC $\beta_s$ and elongation parameters, and in most cases $\delta_c<1 \%$.

\bibliographystyle{apsrev4-1}
\bibliography{new_profile_iopart}

\end{document}